\documentclass[aps,graphicx,twocolumn]{revtex4}%showpacs,

\usepackage{graphicx}

\begin{document}

%\begin{CJK*}{GBK}{song}

\title{Fast and robust quantum control for multimode interactions by using shortcuts to adiabaticity}

\author{Hao Zhang, Xue-Ke Song, Qing Ai, Haibo Wang, Guo-Jian Yang, and Fu-Guo Deng\footnote{Corresponding author: fgdeng@bnu.edu.cn}}

\address{Department of Physics, Applied Optics Beijing Area Major Laboratory, Beijing Normal University, Beijing 100875, China}

\date{\today}

\begin{abstract}
Adiabatic quantum control is a very important approach for quantum physics and quantum information processing. It holds the advantage with robustness to experimental imperfections but accumulates more decoherence due to the long evolution time. Here, we propose a universal protocol for fast and robust quantum control in multimode interactions of a quantum system by using shortcuts to adiabaticity. The results show this protocol can speed up the evolution of a multimode quantum system effectively, and it can also keep the robustness very good while adiabatic quantum control processes cannot. We apply this protocol for the quantum state transfer in quantum information processing in the photon-phonon interactions in an optomechanical system, showing a perfect result. These good features make this protocol have the capability of improving effectively the feasibility of the practical applications of multimode interactions in quantum information processing in experiment.
\end{abstract}

%\pacs{03.67.-a, 03.67.Lx, 42.50.-p, 42.50.Ex}
\maketitle

\section{Introduction}

As a very important approach, adiabatic quantum control has been used widely in quantum physics. Typical applications are the rapid adiabatic passage technique \cite{NVVitanovARPC2001} and the stimulated Raman adiabatic passage technique \cite{KBergmannRMP1998} for two-level and three-level quantum systems, respectively. According to the adiabatic theorem \cite{MBorn1928}, if the state of a quantum system remains non-degenerate and starts in one of the instantaneous eigenstates, it will evolve along this initial state in an adiabatic process. However, the change of coupling, described by a small parameter $\varepsilon$, results in an undesired transition between the different eigenstates with the order of $\exp(-1/\varepsilon)$ \cite{JPDavis1976,JTHwang1977}. Therefore, to suppress the undesired transition, the adiabatic approach should be changed slowly enough but accumulates more total decoherence in practice. Both the intrinsic undesired transition and the environmental decoherence usually lead to the evolution error and decrease the fidelity. Usually a fast process with a high fidelity is needed in quantum information processing, such as quantum computing. To overcome this conflicting, several protocols, called shortcuts to adiabaticity (STA) \cite{ETorronteguiAAMOP2013,HRLewisJMP1969,MDemirplakJPCA2003,MDemirplakJPCB2005,MVBerry2009,XChenPRL20102,XChenPRA2011,AdelCampoPRL2013,SIbPRL2012,SMartinezGaraotPRA2014,ABaksicPRL2016}, have been proposed, such as the transitionless quantum driving (TQD) \cite{MDemirplakJPCA2003,MDemirplakJPCB2005,MVBerry2009,XChenPRL20102,XChenPRA2011,AdelCampoPRL2013}. STA has been used for some practical applications in quantum information processing \cite{MLuPRA2014,MLuLP2014,ACSantosSR2015,YLiangPRA2015,ACSantosPRA2016,SHeSR2016,YHKangSR2016,XKSongNJP2016,XKSongPRA2016,JLWuOE2017,HLMortensenNJP2018},
such as quantum state transfer \cite{MLuLP2014,HLMortensenNJP2018}, entanglement generation \cite{MLuPRA2014,SHeSR2016} and quantum gates \cite{ACSantosSR2015,YLiangPRA2015,ACSantosPRA2016,XKSongNJP2016}.
In recent years, STA has drawn a wide attention and been demonstrated well in experiment in different systems, such as  Bose-Einstein condensates in optical lattices \cite{MGBason2012}, cold atoms \cite{YXDuNC2016}, trapped ions \cite{AnNC2016} and nitrogen-vacancy centers in diamond \cite{JZhangprl2013,BBZhou2016}.

Multimode interactions are common in quantum physics and have been studied in some physical systems. For example, in optomechanical systems \cite{MAspelmeyerRMP2014,Optoadd1,OpomechLongOE,Optoadd2}, photon-phonon-photon \cite{CHDongscience,YDWangPRL2012,LTianPRL2012,LTianPRL2013,YDWangPRL2013} and phonon-photon-phonon \cite{LinQNC,MasselNC,SpethmannNP,Kuzykpra17} interactions are two typical multimode interactions. Effectively controlling multimode interactions is a very important task for many practical applications of a quantum system in quantum information processing, such as quantum state transfer \cite{YDWangPRL2012,LTianPRL2012} and entanglement generation \cite{LTianPRL2013,YDWangPRL2013}. The adiabatic quantum control protocols for multimode interactions have been proposed in optomechanical systems \cite{YDWangPRL2012,LTianPRL2012}. Those protocols also have a big challenge in conflicting between speed and robustness.

To solve the challenge in conflicting between speed and robustness, here, we propose the first protocol to realize a fast and robust quantum control for multimode interactions of a quantum system. It has the following remarkable advantages. First, it is faster and can improve the speed of the whole multimode interaction process, which decreases the total decoherence of the quantum system in quantum information processing. Second, compared to the adiabatic process, it is transitionless and has a higher fidelity. Third, it is robust to the experimental imperfection of time interval. These good features make it improve effectively the feasibility of practical applications of multimode interactions in quantum information processing and quantum state engineering.

This article is organized as follows:  We first introduce multimode interactions and its adiabatic quantum control in Sec.~\ref{sec2}. In Sec.~\ref{sec3}, we calculate the fast and robust quantum control protocol for multimode interactions by using STA and compare it with adiabatic approach. For physical implementation, we perform the fast protocol for photon-phonon interaction in optomechanical system in Sec.~\ref{sec4} and analyze its robustness and fidelity in Sec.~\ref{sec5}. A conclusion is given in Sec.~\ref{sec4}. In Appendix \ref{TQDmultimode}, we give the transitionless quantum driving algorithm for multimode interactions.

\begin{figure}[!ht]
\begin{center}
\includegraphics[width=6.0cm,angle=0]{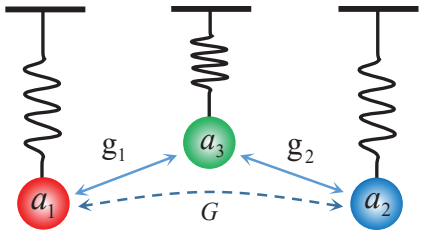}
\caption{Schematic diagram for universal multimode interactions in a quantum system. Each mode can be represented by a harmonic oscillator. $a_{3}$ is a intermediate mode for connecting two others.}\label{basicmodel}
\end{center}
\end{figure}

\section{Multimode interactions and adiabatic quantum control} \label{sec2}

We show the three-mode quantum system in Fig.~\ref{basicmodel}. Two harmonic oscillators couple to each other via the intermediate one. The effective Hamiltonian of this system is given by
\begin{eqnarray}      \label{eq0}
\hat{H}_{1}=\text{g}_{1}\hat{a}_{1}^{\dag}\hat{a}_{3}+\text{g}_{2}\hat{a}_{2}^{\dag}\hat{a}_{3}+\text{H.c.},
\end{eqnarray}
where $\hat{a}_{i}$$(\hat{a}^{\dag}_{i})$ $(i=1,2,3)$ are the annihilation (creation) operators for the corresponding $i$-th mode with the frequency $\omega_{i}$, respectively. $\text{g}_{i}(t)$ is the effective coupling strength between the corresponding mode with the intermediate one. The Heisenberg equations of the system can be derived as
\begin{eqnarray}      \label{eq4}
id\vec{v}(t)/dt=M(t)\vec{v}(t),
\end{eqnarray}
where the vector operator
$\vec{v}(t)=[\hat{a}_{1}(t),\hat{a}_{3}(t),\hat{a}_{2}(t)]^{T}$, and the matrix $M(t)$
can be expressed as
\begin{eqnarray}      \label{eqM}
M(t)=
\left[
\begin{array}{ccc}
0&\text{g}_{1}(t)&0\\
\text{g}_{1}(t)&0&\text{g}_{2}(t)\\
0&\text{g}_{2}(t)&0\\
\end{array}
\right].
\end{eqnarray}
The matrix $M(t)$ has the eigenvalues $\lambda=0,\pm \text{g}_{0}$ and the
corresponding eigenmodes are $\psi_{1}=[-\text{g}_{2}/\text{g}_{0},0,\text{g}_{1}/\text{g}_{0}]^{T}$ and $\psi_{2,3}=[\text{g}_{1}/\text{g}_{0},\pm1,\text{g}_{2}/\text{g}_{0}]^{T}/\sqrt{2}$, where $\text{g}_{0}=\sqrt{\text{g}^{2}_{1}+\text{g}^{2}_{2}}$. The eigenmode $\psi_{1}$ with the eigenvalues $\lambda=0$ is a dark mode which decouples with the intermediate mode.

The adiabatic quantum control is a very effective approach for quantum information processing between the modes 1 and 2, such as the quantum state transfer between these two modes. It is robust against the dissipation from the intermediate mode by evolving along the dark state adiabatically. However, to suppress the undesired transition between different eigenmodes, the near perfect adiabatic process needs so long time that accumulates more decoherence in the whole process. This infidelity is a big challenge for an adiabatic quantum control.

\section{Fast and robust quantum control for multimode interactions}\label{sec3}

\begin{figure*}[ht]
\begin{center}
\includegraphics[width=16cm,angle=0]{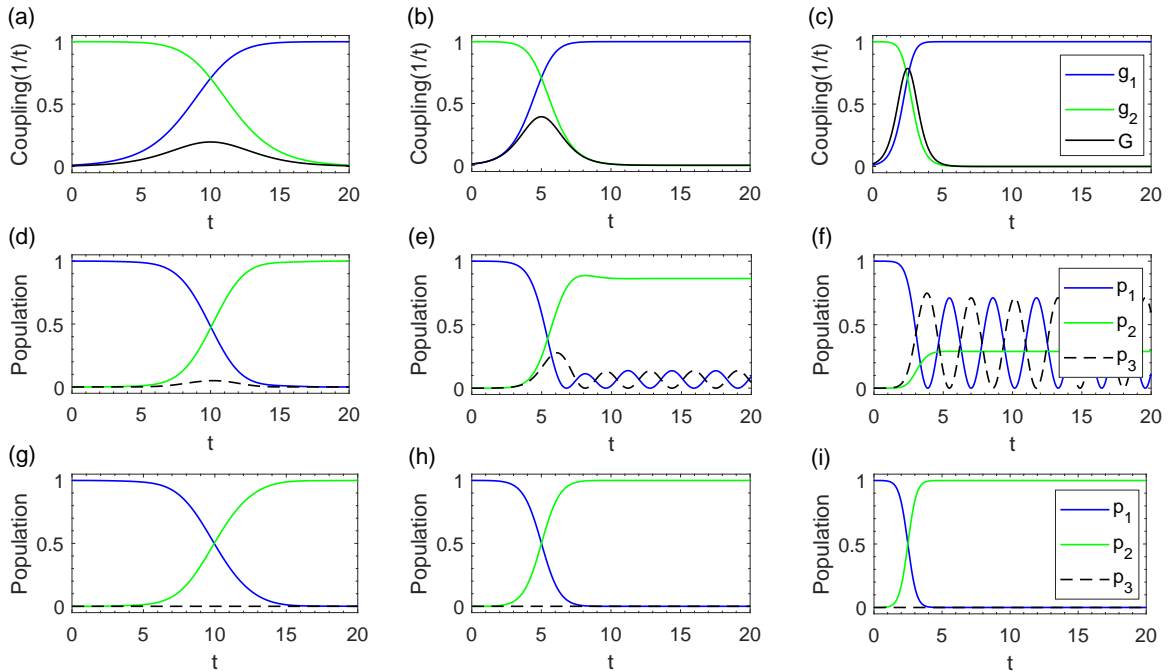}
\caption{ Simulation of two quantum control approaches for the quantum state transfer in multimode interactions. The initial state is prepared in mode $a_{1}$ with the Fock state $|1\rangle$. All the labels for curves with different colors are given in the last figures of every row. $\text{p}_{1}$, $\text{p}_{2}$ and $\text{p}_{3}$ are the populations for the modes $a_{1}$, $a_{2}$ and $a_{3}$, respectively.  (a)-(c) are the variation of coupling strengths. (d)-(f) are the corresponding adiabatic quantum control processes. (g)-(i) are the corresponding fast and robust quantum control processes. The parameter changes as follow: (a) $\nu=0.5$ MHz; (b) $\nu=1$ MHz; (c) $\nu=2$ MHz.}\label{simulation3modes}
\end{center}
\end{figure*}

To improve the speed and mitigate the infidelity, we use the STA to speed up the adiabatic process in three-mode interactions of a quantum system. The main obstacle for accelerating the adiabatic process is the transition amplitude between eigenmodes which will increase when the speed is accelerated. If the speed is up, the evolution path will deviate the original way and more inherent errors are produced. Therefore, the robustness becomes lower. To overcome this mutual restriction, removing the transition is the first task.
According to the TQD algorithm \cite{MVBerry2009}, we can find a new way with no undesired transitions and calculate the matrix $M_{1}(t)$ by using the formula (see appendix for more details) given by
\begin{eqnarray}        \label{eqM1}
M_{1}(t)=i\sum_{n}|\frac{\partial \psi_{n}(t)}{\partial t}\rangle\langle \psi_{n}(t)|,
\end{eqnarray}
where $|n(t)\rangle$ is the eigenmode. Substituting the all
eigenmodes into Eq. (\ref{eqM1}), one can get a new matrix $M'(t)$ given by
\begin{eqnarray}      \label{eqM2}
M_{1}(t)=i\sum_{n=1}^{3} \dot{\psi}_{n}\psi_{n}^{\dag}
=
\left[
\begin{array}{ccc}
0&0&iG\\
0&0&0\\
-iG&0&0\\
\end{array}
\right],
\end{eqnarray}
where $G=(\dot{\text{g}}_{1}\text{g}_{2}-\text{g}_{1}\dot{\text{g}}_{2})/\text{g}^{2}_{0}$. The result
of the matrix $M_{1}(t)$ indicates that there exists a direct
transition between the modes 1 and 2. We choose the adiabatic coupling strengths with `Vitanov' shape \cite{Vasilevpra} given by
\begin{eqnarray}        \label{eqVitanov}
&\text{g}_{1}(t)=\text{g}_{0}\sin(\theta(t)),\nonumber\\
&\text{g}_{2}(t)=\text{g}_{0}\cos(\theta(t)),\nonumber\\
&\theta(t)=\frac{\pi}{2}\frac{1}{1+e^{-\nu(t-5/\nu)}}.
\end{eqnarray}
Here, $\nu$ describes the duration of coupling strengths. Therefore, with above functions, the transitionless matrix $M_{1}$ is given by making $G=\dot{\theta}(t)$ in Eq.(\ref{eqM2}). The comparison between the adiabatic quantum control and the accelerated TQD protocol in quantum state transfer process is shown in Fig.~\ref{simulation3modes}. We change the duration of coupling strengths and the adiabatic processes are accelerated. But the results become imperfect. However, by using the TQD, the new quantum control processes always keep the perfect result in the population transfer. Comparing (g), (h) and (i) in Fig.~\ref{simulation3modes}, the final time for completing the quantum state transfer becomes shorter. The total time in Fig.~\ref{simulation3modes} (i) is three times shorter than Fig.~\ref{simulation3modes} (g).\\

\begin{figure}[!ht]
\begin{center}
\includegraphics[width=8.0cm,angle=0]{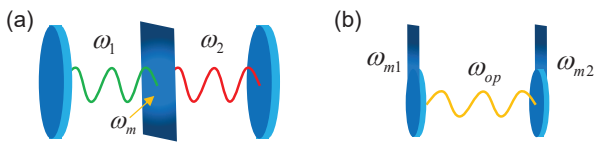}
\caption{(a) Practical schematic diagram for photon-phonon-photon interactions induced by the optomechanical system composed of two cavities and a membrane. The two cavity walls and the middle membrane are fixed, but the membrane can vibrate. (b) Schematic diagram for phonon-phonon-phonon  interactions. Two cavity walls are fixed on cantilevers. }\label{figoptomech}
\end{center}
\end{figure}

\begin{figure*}[!ht]
\begin{center}
\includegraphics[width=16.0cm,angle=0]{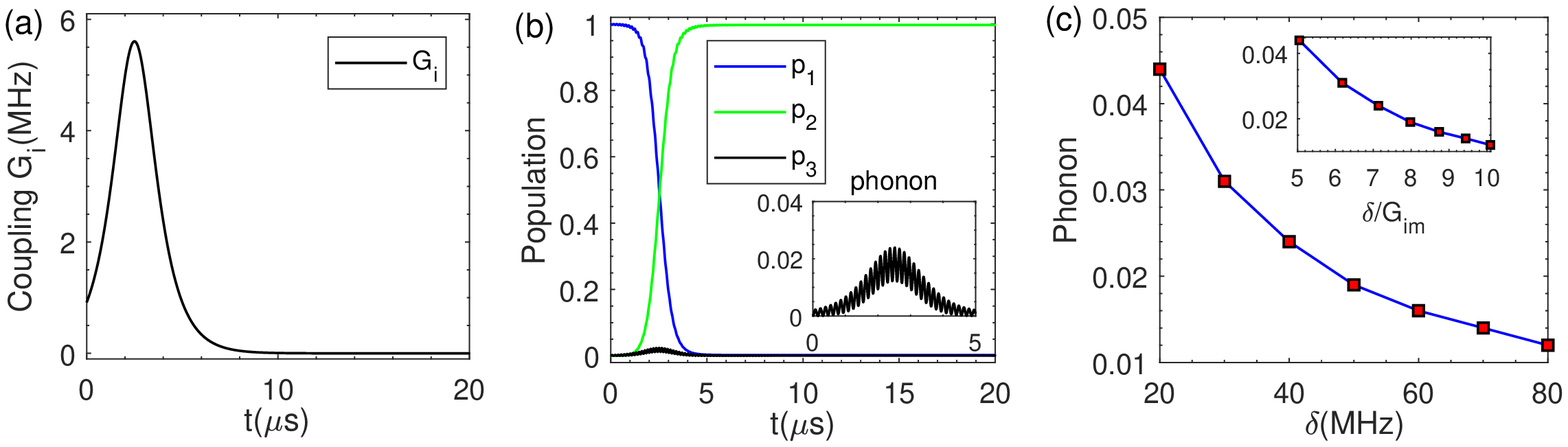}
\caption{ Simulation of the quantum state transfer by using our fast and robust quantum control in optomechanical interactions. (a) The shape of new coupling strengths with the parameters $\nu=2$ MHz and $\delta=40$ MHz. (b) The process of the population transfer by using the coupling in (a). $P_{1}$, $P_{2}$ and $P_{3}$ are the populations for the optical modes 1, 2 and the mechanical mode, respectively. The population of the phonon is amplified in the insert. (c) Variation of the maximal average phonon number with detuning. The insert is the variation of the maximal average phonon number vs the rate between detuning and the maximal coupling strength.}\label{poputqd}
\end{center}
\end{figure*}

\section{Practical application for photon-phonon interactions}\label{sec4}

A typical physical system for photon-phonon interactions is an optomechanical one. For three-mode interactions in this quantum system,
there are two kinds of interactions, i.e., the photon-phonon-photon interaction shown in Fig.~\ref{figoptomech}(a) and the phonon-photon-phonon interaction shown in Fig.~\ref{figoptomech}(b).
Here, we consider Fig.~\ref{figoptomech}(a) in which an optomechanical system composed of two cavity modes coupled to each other via the mechanical mode. Two optical cavities connect to each other via a middle membrane.
After the standard linearization procedure, the Hamiltonian of the system is given by ($\hbar=1$)
\cite{LTianPRL2012,YDWangPRL2012}
\begin{eqnarray}        \label{eqH2}
\hat{H}_{2}=\omega_{m}\hat{b}^{\dag}\hat{b}+\!\!\!\sum_{i=1,2} \!\!\left[-\Delta_{i}\hat{a}^{\dag}_{i}\hat{a}_{i}
+G_{i}(\hat{a}^{\dag}_{i}+\hat{a}_{i})(\hat{b}+\hat{b}^{\dag})\right]\!,
\end{eqnarray}
where $\hat{a}_{i}$$(\hat{a}^{\dag}_{i})$ $(i=1,2)$ and $\hat{b}$$(\hat{b}^{\dag})$ are the annihilation (creation) operators for the $i$-th cavity mode with
frequency $\omega_{i}$ and the mechanical mode, respectively.
$\omega_{m}$ is the mechanical frequency.
$\Delta_{i}=\omega_{di}-\omega_{i}$ and $G_{i}=G_{0i}\sqrt{n_{i}}$
are the laser detuning and the effective linear coupling strength,
respectively. $G_{0i}$ and $n_{i}$ are the single-photon
optomechanical coupling rate and the intracavity photon number induced
by the driving field with the frequency $\omega_{di}$, respectively. We
consider the case that both cavity modes are driven near the red
sidebands. In the interaction picture, the Hamiltonian of the system becomes
(under the rotating-wave approximation)
\begin{eqnarray}      \label{eqH3}
\hat{H}_{3}=\sum_{i=1,2}\delta_{i}\hat{a}_{i}^{\dag}\hat{a}_{i}+G_{i}(\hat{a}_{i}^{\dag}\hat{b}_{m}+\hat{b}_{m}^{\dag}\hat{a}_{i}),
\end{eqnarray}
where $\delta_{i}=-\Delta_{i}-\omega_{m}$. The Heisenberg equation of the system can be derived with
\begin{eqnarray}      \label{eq4}
id\vec{v}_{op}(t)/dt=M_{op}(t)\vec{v}_{op}(t),
\end{eqnarray}
where the vector operator
$\vec{v}_{op}(t)=[\hat{a}_{1}(t),\hat{b}_{m}(t),\hat{a}_{2}(t)]^{T}$, and the matrix $M_{op}(t)$ is the same as Eq.(\ref{eqM}) with the condition $\delta_{i}=0$. There exists an optomechanical dark mode \cite{CHDongscience} and it is used to make the adiabatic control with quantum information processing between two optical modes.

To realize a fast and robust quantum control for above process, we usually need seek a direct coupling between two optical modes based on Eq.(\ref{eqM2}). Here, it is hard to construct a direct coupling and we replace it with an effective two-mode interaction. We set the large detuning condition with $\delta_{i}\gg G_{i}$ in Eq.(\ref{eqH3}) and the large energy offsets suppress the transitions between the optical mode and the mechanical mode \cite{JICiracPRL1997,HKLiPRA2013}. Hence, one can adiabatically eliminate the mechanical mode $\hat{b}$ and obtain the effective beam-splitter-like Hamiltonian
\begin{eqnarray}        \label{eqH4}
\hat{H}_{4}=\sum_{i=1,2}(\delta_{i}+\Omega_{i})\hat{a}_{i}^{\dag}\hat{a}_{i}
+\Omega(\hat{a}^{\dag}_{1}\hat{a}_{2}+\hat{a}^{\dag}_{2}\hat{a}_{1}),
\end{eqnarray}
where $\Omega_{i}=G^{2}_{i}/\delta_{i}$ and
$\Omega=G_{1}G_{2}(\delta^{-1}_{1}+\delta^{-1}_{2})/2$. We set
$\delta_{1}+\Omega_{1}=\delta_{2}+\Omega_{2}$ and
$\delta=\delta_{1}=\delta_{2}$. In the new interaction picture
under the Hamiltonian
$\hat{H}_{0}=\sum_{i=1,2}(\delta_{i}+\Omega_{i})\hat{a}_{i}^{\dag}\hat{a}_{i}$, one
can derive the matrix $M_{2}(t)$ in the Heisenberg picture with
\begin{eqnarray}      \label{eqM3}
M_{2}(t)=
\left[
\begin{array}{ccc}
0&0&\frac{G_{1}G_{2}}{\delta}\\
0&0&0\\
\frac{G_{1}G_{2}}{\delta}&0&0\\
\end{array}
\right].
\end{eqnarray}
The effective matrix $M_{2}(t)$  is equivalent to  $M_{1}(t)$ shown in
Eq. (\ref{eqM1}) derived by the TQD algorithm, when
\begin{eqnarray}        \label{eqGG}
\frac{G_{1}G_{2}}{\delta}=\frac{\text{g}_{1}\dot{\text{g}}_{2}-\dot{\text{g}}_{1}\text{g}_{2}}{\text{g}^{2}_{0}}.
\end{eqnarray}

The evolutions of the photon and the phonon in the system are plotted in Fig.~\ref{poputqd} by using the Hamiltonian in Eq.(\ref{eqH3}).
With the adiabatic coupling Eq.(\ref{eqVitanov}),  Eq.(\ref{eqGG}) becomes $G_{1}G_{2}=\delta\dot{\theta}(t)$.
Here, we design the coupling with $G_{1}=G_{2}$ shown in Fig.~\ref{poputqd}(a).
The other parameters should be chosen to satisfy the large detuning condition $\delta'\gg G_{i}$ and $\delta_{1}+\Omega_{1}=\delta_{2}+\Omega_{2}$. Here we choose $\delta=40$ MHz and $\nu=2$ MHz.

The photon in cavity 1 is transferred to cavity 2 by using our protocol with a higher fidelity, shown in Fig.~\ref{poputqd}(b). The phonon number is suppressed in the whole process and the maximum value is about 0.02 due to the large detuning interaction. From Fig.~\ref{poputqd}(c), one can see that the maximal average phonon number is inversely proportional to the rate between the detuning and the maximal coupling strength.

\begin{figure}[!ht]
\begin{center}
\includegraphics[width=8.0cm,angle=0]{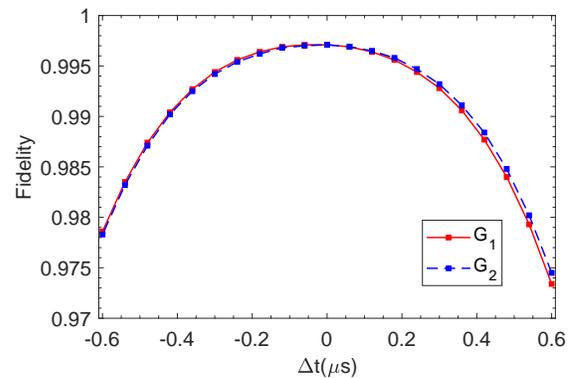}
\caption{Fidelities for the parameter time deviation $\Delta t$. The range of the time interval is $[-0.60,0.60]$. The positive and negative values represent the time delay and advance, respectively. $G_{1}$ ($G_{2}$) is the result from changing $G_{1}$ ($G_{2}$) and keeping  $G_{2}$ ($G_{1}$) unchanged. The parameters are chosen to be $\nu=2$ MHz and $\delta=40$ MHz.}\label{robustness}
\end{center}
\end{figure}

\section{Analysis of robustness and fidelity}\label{sec5}

It is hard to accurately control the time interval between two
coupling strengths in the actual experiment. Therefore we consider
the influence caused by a small deviation of the time interval in
Fig.~\ref{robustness}. We tune the time interval from $-0.6\mu$s to
$0.6\mu$s, and the results indicates that our protocol is
insensitive to the deviation of the time interval.

\begin{figure}[!ht]
\begin{center}
\includegraphics[width=8.0cm,angle=0]{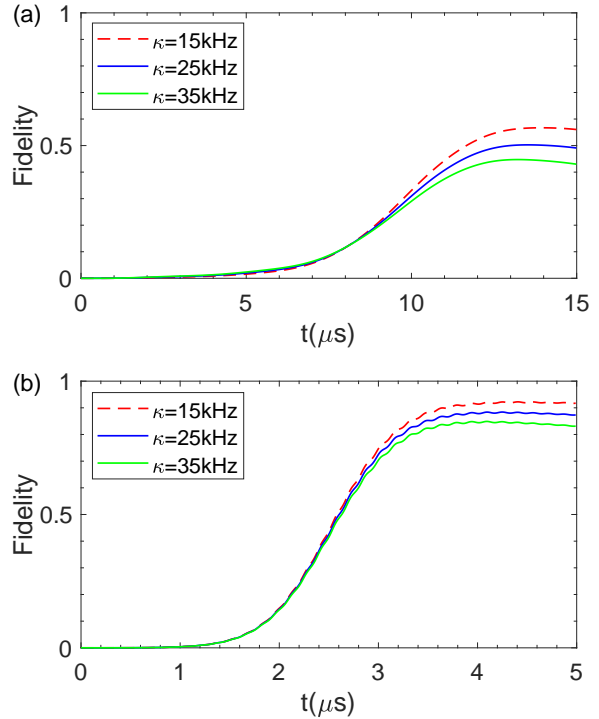}
\caption{Fidelity of the TQD process in the present of different dissipations $\kappa$ in optomechanical interactions. (a) Fidelities for the adiabatic quantum control. Here, we choose $\nu=0.5$ MHz. (b) Fidelities for the fast quantum control process. Parameters are chosen as $\gamma_{m}=500$ Hz and $n_{th}=100$.}\label{fidelity}
\end{center}
\end{figure}

When the dissipation of the mechanical oscillator and the decay of
the cavity is taken into account, the dynamics of the quantum
system described by the Lindblad form master equation is expressed
by
\begin{eqnarray}        \label{eq2}
\frac{d\hat{\rho}}{dt}=i[\hat{\rho},\hat{H}_{s}(t)]\!+\kappa_{1}\hat{L}[\hat{a}_{1}]\hat{\rho}
\!+\kappa_{2}\hat{L}[\hat{a}_{2}]\hat{\rho}\!+\gamma_{m}\hat{D}[\hat{b}_{m}]\hat{\rho},\;\;\;\;
\end{eqnarray}
where $\hat{\rho}$ and $\hat{H}_{s}(t)$ are the density matrix and the
Hamiltonian of the multimode interactions, respectively.
$\kappa_{1}$ and $\kappa_{2}$ represent the decay rates of the cavities 1 and 2,
respectively. $\gamma_{m}$ is the mechanical damping rate.
$\hat{L}[\hat{A}]\hat{\rho}=(2\hat{A}\hat{\rho} \hat{A}^{\dag}-\hat{A}^{\dag}\hat{A}\hat{\rho}-\hat{\rho} \hat{A}^{\dag}\hat{A})/2$.
$\hat{D}[\hat{A}]\hat{\rho}=(n_{th}+1)(2\hat{A}\hat{\rho} \hat{A}^{\dag}-\hat{A}^{\dag}\hat{A}\hat{\rho}-\hat{\rho}
\hat{A}^{\dag}\hat{A})/2+n_{th}(2\hat{A}^{\dag}\hat{\rho} \hat{A}-\hat{A}\hat{A}^{\dag}\hat{\rho}-\hat{\rho}
\hat{A}\hat{A}^{\dag})/2$, where $n_{th}$ is the thermal phonon number of the
environment. Here, we choose the Hamiltonian $\hat{H}_{3}$ to calculate the master equation.
We choose the parameters $\gamma_{m}=500$ Hz and $n_{th}=100$ and change the decay $\kappa=\kappa_{1}=\kappa_{2}$. We calculate the fidelity with the formula $F=\langle
01|tr_{m}[\hat{\rho}(t)]|01\rangle$. Here $|01\rangle$ represents the
state which there are zero and one photon in the cavities 1 and 2,
respectively. $tr_{m}[\hat{\rho}(t)]$ is the reduced density matrix by
tracing the mechanical oscillator degree of freedom. The fidelities are
plotted in Fig.~\ref{fidelity}(a) and (b) for the adiabatic and fast protocols, respectively. The fidelities of the fast protocol are higher than the adiabatic protocol. As the process is accelerated, the total evolution time is shorter.

\section{Conclusion}\label{sec6}

To conclude, we have proposed a universal protocol to realize a fast and robust quantum control for multimode interactions in a quantum system by using STA. We also have applied this protocol in a practical photon-phonon-photon interaction model for the quantum state transfer in quantum information processing. Our protocol has the following remarkable advantages. First, compared with the adiabatic protocol, it is faster as it can improve the speed of a multimode interaction, which reduces more decoherence in the whole evolution process of the quantum system. Second, under the fast process, there is no undesired transition between eigenmodes. Third, it is also robust to some practical experimental imperfections. These good features can effectively improve the feasibility of practical applications of multimode interactions in quantum information processing in experiment.

\section*{ACKNOWLEDGMENT}

We thank Jing Qiu and Xiao-Dong Bai for useful discussion. This work was supported by the National Natural Science Foundation of China (Grants No. 11674033, No. 11474026, No. 11654003, No. 11505007, and No. 61675028).

\section*{APPENDIX: TRANSITIONLESS QUANTUM DRIVING ALGORITHM FOR MULTIMODE INTERACTIONS}\label{TQDmultimode}

Let us consider a quantum system with an arbitrary time-dependent
Hamiltonian $\hat{H}_{0}(t)$. The dynamical process is described by
\begin{eqnarray}      \label{eq4}
\hat{H}_{0}(t)|n(t)\rangle=E_{n}(t)|n(t)\rangle,
\end{eqnarray}
where $|n(t)\rangle$ and $E_{n}(t)$ are the instantaneous eigenstate
and the eigenenergies of $\hat{H}_{0}(t)$, respectively. According
to the adiabatic approximation \cite{MBorn1928}, the dynamical
evolution of states driven by $\hat{H}_{0}(t)$ could be expressed as
\begin{eqnarray}      \label{eq4}
|\psi_{n}(t)\rangle\!\!=\!\exp\left\{\!-i\!\!\int^{t}_{0}\!\!\!dt'E_{n}(t')
\!-\!\!\int^{t}_{0}\!\!\!dt'\langle n(t')|\dot{n}(t')\rangle\right\}|n(t)\rangle,\nonumber\\
\end{eqnarray}
where $|\psi_{n}(t)\rangle$ is the state of system at time $t$. Now,
we seek a new Hamiltonian $\hat{H}(t)$ based on the reverse
engineering approach to satisfy the Schr\"{o}dinger equation
\begin{eqnarray}      \label{eq4}
i|\dot{\psi}_{n}(t)\rangle=\hat{H}(t)|\psi_{n}(t)\rangle.
\end{eqnarray}
Any time-dependent unitary operator $\hat{U}(t)$ is also given by
\begin{eqnarray}      \label{eqU}
i\dot{\hat{U}}(t)=\hat{H}(t)\hat{U}(t),
\end{eqnarray}
and
\begin{eqnarray}      \label{eqH}
\hat{H}(t)=i\dot{\hat{U}}(t)\hat{U}^{\dag}(t).
\end{eqnarray}
In order to guarantee no transitions between the eigenstates of $\hat{H}_{0}(t)$ for any time, the time-dependent unitary operator $\hat{U}(t)$ should have the form
\begin{eqnarray}      \label{eq4}
\hat{U}(t)\!&\!=\!&\!\sum_{n}\exp\left\{-i\!\int^{t}_{0}\!\!dt'E_{n}(t')-\!\!\int^{t}_{0}\!\!dt'\langle n(t')|\dot{n}(t')\rangle\right\}\nonumber\\
&\!\!&\!\otimes|n(t)\rangle\langle n(0)|.
\end{eqnarray}
According to Eq. (\ref{eqH}), the new Hamiltonian is given by
\begin{eqnarray}      \label{eq4}
\hat{H}(t)\!&\!=\!&\!\sum_{n}|n\rangle E_{n}\langle n|+i\sum_{n}(|\dot{n}\rangle\langle n|
-\langle n|\dot{n}\rangle|n\rangle\langle n|)\nonumber\\
&\!=\!&\!\hat{H}_{0}(t)+\hat{H}_{cd}(t),
\end{eqnarray}
where $\hat{H}_{cd}(t)$ is the counter-diabatic
Hamiltonian. The transitions of order $\exp(-1/\varepsilon)$
generated by $\hat{H}_{0}(t)$ can be eliminated by
$\hat{H}_{cd}(t)$. One can find infinitely many Hamiltonians
$\hat{H}(t)$ which differ from each other only by a phase. We
disregard the phase factors
\cite{ETorronteguiAAMOP2013,ACSantosPRA2016} and give the simplest
Hamiltonian  derived with \cite{MVBerry2009}
\begin{eqnarray}      \label{eqsH}
\hat{H}(t)=i\sum_{n}|\dot{n}\rangle\langle n|.
\end{eqnarray}
Here $\hat{H}$ is the couterdiabatic term and purely non-diagonal in the
basis $\{|n\rangle\}$.

In a three-mode quantum system,
we should calculate a new matrix $M'(t)$ which determines the evolution of Heisenberg equations with no undesired transition.
First, we assume the time-dependent evolution operator for  $M'(t)$ which is given by
\begin{eqnarray}        \label{eq1}
i\dot{\hat{U}}_{H}(t)=M'(t)\hat{U}_{H}(t).
\end{eqnarray}
One can solve the $M'(t)$ matrix via the equation given by
\begin{eqnarray}        \label{eqMM}
M'(t)&=&i\left[\frac{\partial}{\partial t}\hat{U}_{H}(t)\right]\hat{U}_{H}^{\dag}(t)\nonumber\\
&=&i\sum_{n}|\frac{\partial \psi_{n}(t)}{\partial t}\rangle\langle \psi_{n}(t)|,
\end{eqnarray}
where $\hat{U}_{H}(t)=\sum_{n}|\psi_{n}(t)\rangle\langle \psi_{n}(0)|$. Here, $|\psi_{n}(t)\rangle$ is the eigenmode of the three-mode quantum system.

%\end{CJK*}

\end{document}